\documentclass[aps,superscriptaddress,preprint]{revtex4}%
\usepackage{mathrsfs}
\usepackage{amsfonts}
\usepackage{amsmath}
\usepackage{amssymb}
\usepackage{graphicx}
\usepackage{diagbox}
\usepackage{booktabs}
\usepackage{epstopdf}
\usepackage{graphicx}
\usepackage{mathrsfs}
\usepackage{bm}
\usepackage{dcolumn}
\usepackage{epstopdf}
\usepackage{dsfont}
\usepackage{tabularx}
\usepackage{float}
\usepackage{color}
\usepackage{pifont}
\usepackage{siunitx}
\usepackage{multirow}%
\providecommand{\U}[1]{\protect\rule{.1in}{.1in}}
\begin{document}
\title{Magnetic ground state of monolayer CeI$_{2}$: Occupation matrix control and
DFT+U calculations}
\author{Yue-Fei Hou}
\affiliation{Institute of Applied Physics and Computational Mathematics, Beijing 100088, China}
\affiliation{Graduate School, China Academy of Engineering Physics, Beijing 100088, China}
\author{Shujing Li}
\affiliation{College of Mathematics and Physics, Beijing University of Chemical Technology,
Beijing 100029, China}
\author{Xinlong Yang}
\affiliation{School of physics, Beijing institute of technology, Beijing 100081, China}
\author{Wei Jiang}
\affiliation{School of physics, Beijing institute of technology, Beijing 100081, China}
\author{Qiuhao Wang}
\affiliation{School of physics, Beijing institute of technology, Beijing 100081, China}
\author{Fawei Zheng}
\affiliation{School of physics, Beijing institute of technology, Beijing 100081, China}
\author{Zhen-Guo Fu}
\thanks{Corresponding author. Email address: fu\_zhenguo@iapcm.ac.cn}
\affiliation{Institute of Applied Physics and Computational Mathematics, Beijing 100088, China}
\affiliation{National Key Laboratory of Computational Physics, Beijing 100088, China}
\author{Ping Zhang}
\thanks{Corresponding author. Email address: zhang\_ping@iapcm.ac.cn}
\affiliation{Institute of Applied Physics and Computational Mathematics, Beijing 100088, China}
\affiliation{National Key Laboratory of Computational Physics, Beijing 100088, China}
\affiliation{School of Physics and Physical Engineering, Qufu Normal University, Qufu
273165, China}

\begin{abstract}
The magnetic ground state is crucial for the applications of the two-dimension
magnets as it decides fundamental magnetic properties of the material, such as
magnetic order, magnetic transition temperature, and low-energy excitation of
the spin waves. However, the simulations for magnetism of local-electron
systems are challenging due to the existence of metastable states. In this
study, occupation matrix control (OMC) and density functional theory plus
Hubbard $U$ calculations are applied to investigate the magnetic ground state
of monolayer CeI$_{2}$. Following the predicted ferromagnetic (FM) order, the
FM ground state and the FM metastable states are identified and found to have
different values of the magnetic parameters. Based on the calculated magnetic
parameters of the FM ground state, the Curie temperature is estimated to be
$128$ K for monolayer CeI$_{2}$. When spin-orbit coupling (SOC) is considered,
the FM ground state is further confirmed to contain both off-plane and
in-plane components of magnetization. SOC is shown to be essential for
reasonably describing not only magnetic anisotropy but also local electronic
orbital state of monolayer CeI$_{2}$.

\end{abstract}
\maketitle

\section{Introduction}

The two-dimension(2D) magnets exhibit magnificent potential in spintronic
devices and magnetic storage devices due to their unique properties and the
nanoscale size. Although Mermin-Wagner theorem \cite{Mermin} denies the
possibility of long-range magnetic order in 2D systems at non-zero
temperature, a series of experimental studies \cite{Jiang2021, Huang2017,
Tian2019, Cai2019, Zhang2019, Banerjee2017, Zhou2019, Bonilla2018, Li2018,
Hara2018, Gong2017, Wang2016, Kim2019, Flem1982, Fei2018, May2019} have
confirmed the presence of magnetic orders in 2D materials. These findings
encourage people to further explore 2D magnetic materials with superior
properties, such as robust dynamic stability, high magnetic transition
temperature, and unique magnetic anisotropy.

The magnetic ground state (GS) of a 2D material is significant. It not only
dominates the behaviour about magnetism at low temperature, but decides the
low-energy excitation of the spin waves at finite temperature. For magnetic
crystals, a comprehensive description of the magnetic GS should contain the
information of two aspects: the specific magnetic structure and the electronic
states of the magnetic sites. For the former part, spin-polarized scanning
tunneling microscope (SP-STM), Ramam spectroscopy, and the observation of
anomalous Hall effect have been adopted to examine the magnetic order of 2D
magnetic insulators and 2D magnetic metals
\cite{Huang2017,Gong2017,May2019,Lee2016,Tian2016,Deng2018}. For the latter
part, x-ray magnetic circular dichroism (XMCD) and x-ray magnetic linear
dichroism(XMLD) are two techniques which help to obtain the local electronic
configuration \cite{Liu2016,Laan1999}. Nonetheless, exploring magnetic ground
states of 2D materials experimentally is still challenging. The limitations
include the acquirement of high quality samples, strict experimental
conditions, and even expensive costs on economy and time. From this point of
view, exploring 2D magnetic materials by using density functional theory (DFT)
calculations is another feasible scheme. The DFT tool gives the prediction of
the magnetic ground state without synthesizing actual samples. It has been
applied to study a series of 2D magnetic materials such as layered $\alpha
$-RuCl$_{3}$, VSe(Te)$_{2}$, CrGe(Si)Te$_{3}$, etc.
\cite{Sarikurt2018,Kim20161,
Sandilands2015,Li2014,Ataca2012,Chen2020,Fuh2016,Pan2014,Kan2014,Lin2017,Lin2016}
For 2D CrI$_{3}$ and CrBr$_{3}$, DFT calculations give quite consistent
prediction of magnetic features with the experiments, including the magnetic
structure and Curie temperature ($T_{\text{C}}$)
\cite{Wang20162,Zhang2015,Liu20162,Jiang20212}. With the assistance of DFT
calculations, the studying progress of 2D magnetic materials has been
accelerated significantly.

In 2D magnetic insulators and semiconductors, the spin magnetic moments are
commonly contributed by polarized electrons from partially filled $d$ or
\textit{f} atomic shells. These electrons are localized around the lattice
sites and need extra corrections for strong electronic correlation. Therefore,
the density functional theory plus Hubbard $U$ (DFT+U) scheme
\cite{Liechtenstein,Dudarev} is widely applied to give reasonable electronic
structures of the correlated systems. In fact, the conventional DFT+U
computational scheme faces challenges in locating the ground state of the
localized electrons due to the presence of metastable states, especially when
the \textit{d} and \textit{f} shells are less filled. In other words, although
a specific magnetic ordered state can be determined by DFT+U calculations,
however, the obtained electronic state of the localized electrons may be not
identical for different researchers. This phenomenon has occurred in the DFT+U
studies of Ti-based systems \cite{Watson}, Ce-based systems
\cite{Watson,Shick}, U-based systems \cite{Miskowiec,Dorado,Dorado1,
Dorado2,Matthew, Claisse}, cubic fluorite PuO$_{2}$ \cite{Jomard,Hou2022}, and
monolayer VX$_{3}$ (X=Cl, Br, I) \cite{hou2024} etc., and the metastable
states can originate from 3$d$, 4$f$, or 5$f$ localized electrons (not found
in 4$d$ or 5$d$ electronic systems yet, to our knowledge). Moreover, these
works have also shown that the occurrence of trapping into metastable states
is not relevant to the form of adopted exchange-correlation functional. Local
density approximation (LDA) \cite{Shick}, generalized gradient approximation
(GGA) \cite{Watson,Miskowiec,Dorado,Dorado1}, strongly constrained appropriate
normed (SCAN) type meta-GGA \cite{Hou2022}, and HSE06 type hybrid
\cite{Jollet,Ratcliff} functionals have been adopted in the studies and this
accident still happens constantly. To solve the problem, a few techniques have
been applied including U-ramping \cite{Claisse,Meredig}, quasi-annealing
\cite{Geng}, occupation matrix control (OMC)
\cite{Watson,Dorado,Claisse,Amadon,Zhou2011,Zhou2022}, DFT+DMFT
\cite{Amadon2012} and even the combination of them \cite{Rabone}. These
techniques help to reduce the probability of trapping into metastable states
to some extent. However, it remains an open question as to how the ground
state of the localized electrons can be strictly obtained.

Recently, the monolayer CeI$_{2}$ was predicted to be an in-plane
ferromagnetic (FM) semiconductor with a high $T_{\text{C}}$ of $374$ K by
DFT+U calculations \cite{Sheng}. The dynamic stability of the monolayer
CeI$_{2}$ was supported by both phonon calculations and a molecular dynamics
simulation at room temperature. In the monolayer CeI$_{2}$, two 6\textit{s}
electrons of the Ce atom move away to form the Ce-I bond, leaving a
$4f^{1}5d^{1}$ electronic configuration for the Ce$^{2+}$ ion. According to
above analysis, a number of metastable states originated from the highly
localized $4f$ electrons should be expected. Owing to the importance for both
basic physics and potential applications, it is necessary to provide a
systematic analysis of the magnetic GS of monolayer CeI$_{2}$.

In this study, we perform DFT+U calculations combined with OMC to investigate
the magnetic GS of monolayer CeI$_{2}$. The magnetic structure is once again
confirmed to be FM. The FM GS as well as twenty-three FM metastable states of
monolayer CeI2 with different local electronic states are identified. Our
results show that the computed isotropy exchange parameter and magnetic
anisotropy energy are distinguishable for the GS and metastable states. Based
on the exchange parameters of the magnetic GS, the $T_{\text{C}}$ is estimated
to be about $128$ K by using Monte Carlo simulations. To reasonably analyze
the contributions to the total energies, the Coulomb potential energies of the
4\textit{f} electron with different occupied electronic orbitals in the
crystal field are discussed. Furthermore, we take spin-orbit coupling (SOC)
into consideration. The magnetic structure of monolayer CeI$_{2}$ then becomes
FM with both in-plane and off-plane components of magnetic moments and the
easy axis of magnetization is coupled with the crystal structure. The single
Ce-$4f$ electron occupies another distinguishable electronic orbital compared
to the situation without SOC, which illustrates that SOC is essential to
describe the magnetic GS of monolayer CeI$_{2}$. A fully off-plane FM state is
confirmed to be a metastable state of monolayer CeI$_{2}$ whose total energy
is just slightly higher than the magnetic GS. We anticipate that achieving
saturation magnetization in the off-plane direction for monolayer CeI$_{2}$ is
feasible under appropriate external conditions.

\section{Computational Methods}

The DFT+U calculations are carried out by employing the Vienna ab-initio
simulation package (VASP) code \cite{Kresse}. OMC is implemented by applying
the patch of Allen \textit{et al}. tailored for VASP \cite{Watson}. The
adopted atomic pseudo potentials are constructed by the project augmented wave
(PAW) method \cite{Blochl,Kresse2,Blochl2}. The cut-off energy of the plane
wave bases is set to $600$ eV. The first Brillouin Zone is sampled by a
$\Gamma$-center $11\times11\times1$ $k$-point mesh. The convergence of both
cut-off energy and $k$-point mesh has been carefully examined. The crystal
structures are relaxed until the Feynman-Hellman force on each atom is smaller
than $0.005$ eV/\r{A}.\ The optimizing of the charge density is ended when the
energy difference is smaller than $10^{-6}$ eV between the current two steps
of the iteration. We make use of the DFT+U method of Dudarev \textit{et al}.
\cite{Dudarev} with a form of%
\begin{equation}
E_{\text{DFT+U}}=E_{\text{DFT}}+\frac{U-J}{2}\sum_{\sigma}\left[  \left(
\sum_{m_{1}}n_{m_{1},m_{1}}^{\sigma}\right)  -\left(  \sum_{m_{1},m_{2}%
}n_{m_{1},m_{2}}^{\sigma}n_{m_{2},m_{1}}^{\sigma}\right)  \right]  ,
\end{equation}
where $n_{m_{1},m_{2}}^{\sigma}$ is the matrix element of the occupation
matrix with spin $\mathit{\sigma}$. The standard DFT part here is handled by
the Perdew-Burke-Ernzerhof (PBE) type GGA exchange-correlation functional.
\cite{Perdew} The Hubbard $U$ and Hund $J$ for the Ce-$4f$ electron are set to
be $7.47$ eV and $0.99$ eV, respectively, which have been applied in previous
study \cite{Sheng, Larson}. To be cautious, a discussion is also made to
demonstrate the rationality of the $U$ parameter (see Section A of the
Supporting Information \cite{SI}). Based on the computed Bloch states with
plane waves bases, Wannier90 code \cite{Mostofi} is used to construct the
tight-binding Hamiltonian. TB2J code \cite{He} is used to calculate the
Green's function and then the isotropic exchange parameter for magnetic
properties is obtained. The $T_{\text{C}}$ is gained by Monte Carlo
simulations based on the heat bath algorithms \cite{Miyatake}. The
visualizations of the crystal structure and spin densities in this work is
achieved by VESTA \cite{Momma}. The VASPKIT code is used to process the data
of the DFT calculations \cite{Wang2021}.

\section{OMC of the $4f^{1}$ electronic state}

A quantum state of a single electron can be express as a linear combination of
a group of complete and orthogonal bases in the Hilbert space. In the
limitation of localized electrons of an isolated atom, the dimension of the
space is equal to $2l+1$, where $l$ is the angular quantum number of that
electron. In monolayer CeI$_{2}$, $l$ is no longer a good quantum number for
the $5d$ electron of Ce$^{2+}$ since the $5d$ electron is not well localized.
Additionally, the delocalized $5d$ electron has a more plane-wave-like charge
density rather than the atom-orbital-like charge density of the $4f$ electron.
Hence only the $4f$ electron is accompanied by high risks of falling into
metastable states in DFT+U calculations. Its electronic orbital requires
careful control and observation.

The occupation matrix is the density matrix in particle-number representation
and a $14\times14$ occupation matrix is sufficient to represent the electronic
state of the single $4f$ electron exactly. Here the general set of 4\textit{f}
atomic orbitals is used as the bases, which are $f_{y(3x^{2}-y^{2})^{\uparrow
}}$, $f_{xyz^{\uparrow}}$, $f_{yz^{2\uparrow}}$, $f_{z^{3\uparrow}}$,
$f_{xz^{2\uparrow}}$, $f_{z(x^{2}-y^{2})^{\uparrow}}$, $f_{x(x^{2}%
-3y^{2})^{\uparrow}}$, $f_{y(3x^{2}-y^{2})^{\downarrow}}$, $f_{xyz^{\downarrow
}}$, $f_{yz^{2\downarrow}}$, $f_{z^{3\downarrow}}$, $f_{xz^{2\downarrow}}$,
$f_{z(x^{2}-y^{2})^{\downarrow}}$, and $f_{x(x^{2}-3y^{2})^{\downarrow}}$,
respectively. The spin-quantization axis is chosen to be along $z$ direction
of the Cartesian coordinate system. We firstly do not take SOC into
consideration. Since there is only one localized electron in the $4f$ shell of
Ce$^{2+}$, any quantum state of the single electron with both spin up and spin
down components is unphysical. We thus neglect the spin index and a quantum
state of the single $4f$ electron can be written as%
\begin{equation}
\psi_{4f}=c_{1}f_{y(3x^{2}-y^{2})}+c_{2}f_{xyz}+c_{3}f_{yz^{2}}+c_{4}f_{z^{2}%
}+c_{5}f_{xz^{2}}+c_{6}f_{z(x^{2}-y^{2})}+c_{7}f_{x\left(  x^{2}%
-3y^{2}\right)  },
\end{equation}
where $c_{i}$ ($i=1,2,\cdots,7$) is the complex expansion coefficient of the
basis. There should be
\begin{equation}
\sum_{i}c_{i}^{2}=1.
\end{equation}
For simplicity, we mark the quantum state as ($c_{1}$, $c_{2}$, $c_{3}$,
$c_{4}$, $c_{5}$, $c_{6}$, $c_{7}$) for the following part of this paper. The
eigenvectors and eigenvalues of the occupation matrix are actually quantum
states and the corresponding occupation numbers, respectively.

An OMC procedure is to set an initial occupation matrix artificially for the
localized electrons in the DFT+U calculations. The initial occupation matrix
has also been called the starting point \cite{Zhou2022}. Different starting
points may access to different final states of the localized electrons. In our
calculations, we make our starting points for the Ce-$4f$ electron all
diagonal with a average occupation on several orbital bases. The down spin
block and the two off-diagonal blocks of the $14\times14$ occupation matrix
are null matrices. The up spin block of the occupation matrix has a form of%
\begin{equation}
\left(
\begin{array}
[c]{cccc}%
a_{1} & 0 & \cdots & 0\\
0 & a_{2} & \cdots & 0\\
\cdots & \cdots & \cdots & \cdots\\
0 & 0 & \cdots & a_{7}%
\end{array}
\right)
\end{equation}
with%
\begin{equation}
a_{i}=0\text{, or }\frac{1}{n}\text{ (}i,n=1,2,\cdots,7\text{),}%
\end{equation}
where $n$ is the number of orbital bases whose occupation number is non-zero.
The trace of the up spin block is the number of occupying electrons, which is
equal to $1$ for the Ce-$4f$ shell here. Hence there is%
\begin{equation}
\sum_{i}a_{i}=1.
\end{equation}
Based on Eqs. (4)-(6), we have $\sum\nolimits_{i=1}^{7}C_{7}^{i}=127$
different starting points for the DFT+U calculations. After the
self-consistent iteration processes based on different starting points, we
obtain twenty-four final states of monolayer CeI$_{2}$, whose $4f$ electronic
states are different with each other. A detailed discussion of the OMC
procedure for the SOC situation is provided in Section B of the Supporting
Information. \cite{SI}

\section{Magnetic properties of the GS and metastable states}

The optimized crystal structure of monolayer CeI$_{2}$ is shown in Fig. 1(a).
Each Ce$^{2+}$ ion is surrounded by six nearest I$^{-}$ ions which form a
regular triangle prism (RTP). The distance between two nearest Ce$^{2+}$ ions
is $4.28$ \AA \ which makes the direct exchange from different Ce sites
negligible. The main reason that causes the magnetic order should be the
superexchange interaction mediated by I-$p$ electrons. Since both Ce-$4f$ and
Ce-$5d$ electrons contribute to the superexchange interaction, the total
inter-site exchange strength can be decomposed to $J^{d-d}$, $J^{d-f}$, and
$J^{f-f}$, as shown in Fig. 1(b). These are the exchange parameter between
$5d$ shell and $5d$ shell, $5d$ shell and $4f$ shell, and $4f$ shell and $4f$
shell from different sites, respectively. Meanwhile, with the help of
coordinated I$^{-}$ ions, the $d-f$ hybridization on each Ce site should be
considered. This on-site $d-f$ exchange parameter is denoted by
$J^{\text{site}}$, as shown in Fig. 1(b). For the inter-site exchange
parameters, we only consider the first nearest exchange parameter $J_{1}$ and
second nearest exchange parameter $J_{2}$ as shown in Fig. 1(a). Hence the
total Hamiltonian of CeI$_{2}$ can be express as
\begin{equation}
\hat{H}_{\text{tot}}=\delta-J_{1}\sum_{\left\langle i,j\right\rangle }\vec
{S}_{i}\cdot\vec{S}_{j}-J_{2}\sum_{\left\langle \left\langle i,j\right\rangle
\right\rangle }\vec{S}_{i}\cdot\vec{S}_{j}-J^{\text{site}}\sum_{i}\vec{S}%
_{i}^{d}\cdot\vec{S}_{i}^{f}%
\end{equation}
with%
\begin{equation}
J_{1}=J^{d-d}+J^{d-f}+J^{f-f},
\end{equation}
where $\mathit{\delta}$ is the spin-independent part of the total energy. $i$
and $j$ denote different sites in the unit cell.

All the following discussions for monolayer CeI$_{2}$ are based on the fact
that the magnetic structure of the GS is FM no matter which orbital the
localized electron is occupying. This feature of CeI$_{2}$ can be verified by
our computed total energies of the magnetic structures, as well as the
computed values of the exchange parameters, which will be discussed later.
With OMC applied, the relative energies of the GS and the metastable states of
FM CeI$_{2}$ are listed in Table I. It is noteworthy that the FM state studied
by the previous study \cite{Sheng} is confirmed to be a metastable state
(S$_{20}$) in our work.

By using a model Hamiltonian of spin interactions like Eq. (7) in combination
with the DFT total energies of different magnetic structures, one can fit the
values of the exchange parameters. However, this scheme is invalid for
monolayer CeI$_{2}$ because the exchange parameter is not identical for
different magnetic structures. A detailed discussion about it is given in
Section C of the Supporting Information. \cite{SI} Here, we make use of the
obtained Bloch states by a self-consistent calculation to construct
tight-binding Hamiltonian and calculate the exchange parameter by using
Green's function method \cite{He}. The values of $J_{1}$ and $J_{2}$ for
monolayer CeI$_{2}$ with different local electronic states are shown in Fig.
1(c). One can find that $J_{1}$ is always positive and $J_{2}$ is always
negative for all the states. The absolute value of $J_{1}$ is far greater than
$J_{2}$, which favors FM order of monolayer CeI$_{2}$. Note that $J_{1}$ is
quite sensitive to the $4f$ electronic states and S$_{0}$ has the largest
$J_{1}$.

The exchange parameters of the ground state S$_{0}$ are used to simulate the
$T_{\text{C}}$ of monolayer CeI$_{2}$. There are $J_{1}=10.7$ meV and
$J_{2}=-0.34$ meV. Monte-Carlo method is used and $T_{\text{C}}$ is estimated
to be about $128$ K as shown in Fig. 1(d). It is important to note that the
presence of metastable states of monolayer CeI$_{2}$ can lead to different
values of the inter-site exchange parameters, potentially affecting the
reliability of the predicted $T_{\text{C}}$ for the material. Therefore, it is
also recommended to exclude the metastable states before conducting further
calculations for other local-electron systems, in addition to monolayer
CeI$_{2}$.

\begin{table}[ptb]
\caption{The 1$^{\text{st}}$ column lists the GS (S$_{0}$) and the metastable
states (S$_{1}$-S$_{23}$). The 2$^{\text{nd}}$ column gives the DFT total
energies of all the FM states relative to the GS denoted as $\Delta E$ (in
unit of meV). The 3$^{\text{rd}}$ column shows the specific representations of
the calculated $4f^{1}$ states with $f_{y(3x^{2}-y^{2})}$, $f_{xyz}$,
$f_{yz^{2}}$, $f_{z^{3}}$, $f_{xz^{2}}$, $f_{z(x^{2}-y^{2})}$, and
$f_{x(x^{2}-3y^{2})}$ as bases. The 4$^{\text{th}}$ to the 6$^{\text{th}}$
columns give the relative energies of $x$, $y$, and $z$ directions of spin
polarization for each state (in unit of $\mu$eV). The 7$^{\text{th}}$ to the
10$^{\text{th}}$ columns give the values of $J^{\text{site}}$, $J^{d-d}$,
$J^{d-f}$, and $J^{f-f}$ for each state (in unit of meV). }%
\label{tableI}
\begin{tabular*}
{\hsize}[c]{@{\extracolsep{\fill}}lccccccccc}\hline\hline
& $\Delta E$ & $4f^{1}$ electronic state & $x$ & $y$ & $z$ & $J^{\text{site}}$
& $J^{d-d}$ & $J^{d-f}$ & $J^{f-f}$\\\hline
\multicolumn{1}{c}{S$_{0}$} & $0$ & $(0.66,0,0,-0.74,0,0,0)$ & $0$ & $0$ &
$393$ & $116.1$ & $10.2$ & $0.30$ & $0.22$\\
\multicolumn{1}{c}{S$_{1}$} & $4.4$ & $(0,0,0,-0.74,0,0,0.66)$ & $0$ & $0$ &
$381$ & $113.9$ & $9.18$ & $0.38$ & $0.28$\\
\multicolumn{1}{c}{S$_{2}$} & $5.0$ & $(0.62,0,0.4,0.6,0,-0.3,0)$ & $0$ & $0$
& $409$ & $104.3$ & $9.10$ & $0.49$ & $0.17$\\
\multicolumn{1}{c}{S$_{3}$} & $5.0$ & $(0.38,0.1,0.33,-0.61,0.29,0,-0.53)$ &
$0$ & $6$ & $406$ & $107.4$ & $8.60$ & $1.42$ & $0.10$\\
\multicolumn{1}{c}{S$_{4}$} & $6.6$ & $(0.62,0,-0.79,0,0,0,0)$ & $0$ & $24$ &
$461$ & $97.3$ & $9.04$ & $0.56$ & $0.24$\\
\multicolumn{1}{c}{S$_{5}$} & $6.6$ & $(0.61,0,0.38,0,-0.69,0,0)$ & $9$ & $0$
& $455$ & $97.3$ & $9.58$ & $0.20$ & $0.03$\\
\multicolumn{1}{c}{S$_{6}$} & $7.3$ & $(0.54,0,0.28,0,0.73,0,0.29)$ & $27$ &
$0$ & $460$ & $96.9$ & $9.22$ & $0.40$ & $0.21$\\
\multicolumn{1}{c}{S$_{7}$} & $7.6$ & $(0.5,0,-0.77,0,0.16,0,-0.34)$ & $0$ &
$25$ & $458$ & $96.7$ & $9.33$ & $0.50$ & $0.10$\\
\multicolumn{1}{c}{S$_{8}$} & $7.6$ & $(0.24,0.1,0.5,-0.45,0.37,0,-0.58)$ &
$0$ & $13$ & $425$ & $101.6$ & $8.78$ & $1.14$ & $0.15$\\
\multicolumn{1}{c}{S$_{9}$} & $8.3$ & $(0.43,0,-0.76,0,0.2,0,-0.43)$ & $0$ &
$25$ & $456$ & $96.3$ & $9.22$ & $0.70$ & $0.01$\\
\multicolumn{1}{c}{S$_{10}$} & $8.4$ & $(0.42,0,0.19,0,-0.76,0,-0.45)$ & $17$
& $0$ & $462$ & $96.2$ & $8.34$ & $1.24$ & $0.13$\\
\multicolumn{1}{c}{S$_{11}$} & $8.5$ & $(0.4,0,0.57,0,0.54,0,-0.47)$ & $0$ &
$12$ & $449$ & $96.2$ & $8.70$ & $1.32$ & $0.18$\\
\multicolumn{1}{c}{S$_{12}$} & $9.9$ & $(0,0,0,0,0.79,0,0.61)$ & $46$ & $0$ &
$460$ & $95.4$ & $8.86$ & $0.67$ & $0.51$\\
\multicolumn{1}{c}{S$_{13}$} & $9.9$ & $(0,0,-0.68,0,-0.4,0,0.61)$ & $0$ &
$23$ & $450$ & $95.4$ & $9.16$ & $0.60$ & $0.22$\\
\multicolumn{1}{c}{S$_{14}$} & $24.1$ & $(0.33,0,0.6,0,-0.52,0,0.52)$ & $0$ &
$15$ & $449$ & $80.8$ & $8.25$ & $1.49$ & $0.38$\\
\multicolumn{1}{c}{S$_{15}$} & $33.3$ & $(0,1,0,0,0,0,0)$ & $44$ & $0$ & $428$
& $49.0$ & $7.83$ & $0.50$ & $0.24$\\
\multicolumn{1}{c}{S$_{16}$} & $33.3$ & $(0,0,0,0,0,1,0)$ & $0$ & $43$ & $427$
& $49.0$ & $7.99$ & $0.60$ & $0.26$\\
\multicolumn{1}{c}{S$_{17}$} & $33.3$ & $(0,-0.92,0,0,0,0.4,0)$ & $29$ & $0$ &
$421$ & $49.0$ & $7.14$ & $1.38$ & $0.62$\\
\multicolumn{1}{c}{S$_{18}$} & $188.9$ & $(0.8,0,0.6,0,0,0,0)$ & $0$ & $144$ &
$363$ & $103.2$ & $9.50$ & $0.46$ & $0.25$\\
\multicolumn{1}{c}{S$_{19}$} & $190.1$ & $(0,0,0,-1,0,0,0)$ & $0$ & $0$ &
$619$ & $109.9$ & $6.86$ & $1.19$ & $0.02$\\
\multicolumn{1}{c}{S$_{20}$} & $267.5$ & $(1,0,0,0,0,0,0)$ & $0$ & $0$ & $101$
& $114.8$ & $9.92$ & $0.38$ & $0.10$\\
\multicolumn{1}{c}{S$_{21}$} & $276.0$ & $(0,0,0,0,0,0,1)$ & $0$ & $0$ & $75$
& $111.3$ & $10.0$ & $0.34$ & $-0.03$\\
\multicolumn{1}{c}{S$_{22}$} & $279.2$ & $(0.25,0,0.97,0,0,0,0)$ & $0$ & $183$
& $698$ & $89.8$ & $7.73$ & $1.40$ & $0.03$\\
\multicolumn{1}{c}{S$_{23}$} & $280.2$ & $(0,0,0,0,0.97,0,-0.25)$ & $175$ &
$0$ & $692$ & $88.5$ & $6.89$ & $1.14$ & $0.36$\\\hline\hline
\end{tabular*}
\end{table}

In the case of weak SOC (common in $3d$ electronic systems), the orbital state
of localized electrons is primarily decided by on-site Coulomb interactions
and the crystal field. Thus, SOC can be treat as an additional perturbation.
In other words, the SOC energy can be safely gained based on the simulated
charge density which does not include SOC effect. The SOC Hamiltonian is
written as
\begin{equation}
H_{\text{SOC}}=-\lambda\vec{L}\cdot\vec{S},
\end{equation}
where $\mathit{\lambda}$ parameter denotes the strength of SOC. Applying this
approximation to FM monolayer CeI$_{2}$, the relative energies of different
spin directions are computed. As shown in Table I, all the magnetic states
have the highest energy when the spin is along $z$ direction. The easy axis of
spin polarization varies depending on the specific orbital state of the
4\textit{f} electron. For S$_{0}$ (the GS), both $x$ and $y$ directions are
the easy axes which are $390$ $\mu$eV lower in energy than the $z$ direction.
Hence, monolayer CeI$_{2}$ is recognized as an isotropic in-plane FM material
within the theoretical framework of weak SOC.

In order to better understand the origin of the FM order in monolayer
CeI$_{2}$, the components of $J_{1}$ are computed and listed in Table I. It is
observed that the values of $J^{d-d}$, $J^{d-f}$, and $J^{f-f}$ are always in
the order of $J^{d-d}>J^{d-f}>J^{f-f}$ for different states. $J_{1}$ is
primarily contributed by $J^{d-d}$, which is similar to the situation in
GdI$_{2}$ (Gd$^{2+}$: $4f^{7}5d^{1}$) \cite{Wang2020}. According to Eq. (7),
$J^{\text{site}}$ can be obtained from the energy difference of the two spin
configurations $4f^{\uparrow}5d^{\uparrow}$ and $4f^{\uparrow}5d^{\downarrow}$
for each state, which is denoted as $J^{\text{site}}=1/2\left(  E_{\text{tot}%
}^{\uparrow\uparrow}-E_{\text{tot}}^{\uparrow\downarrow}\right)  $. The
computed values of $J^{\text{site}}$ for different states are listed in Table
I. The positive values of $J^{\text{site}}$ helps to stabilize the saturated
spin moment of Ce$^{2+}$, i.e. $4f^{\uparrow}5d^{\uparrow}$ spin configuration
is favorable in energy for monolayer CeI$_{2}$ rather than $4f^{\uparrow
}5d^{\downarrow}$ spin configuration, although the spin singlet state ($^{1}%
$G$_{0}^{4}$) has been recommended by laser spectroscopy for Ce atoms
\cite{Worden}. The magnetic moment is 2 $\mu_{B}$ per Ce coming from the two
parallel spins. Furthermore, S$_{0}$ has the largest value of $J^{\text{site}%
}$, which is one of the reason why it leads to the lowest total energy
referring to Eq. (7).

The spin densities and electronic structures of S$_{20}$ and S$_{0}$ are
illustrated in Fig. 2. The significant difference between the two states can
be noticed by inspecting the spin densities. Meanwhile, S$_{0}$ exhibits more
low-energy contributions to the bands from $4f$ orbitals than that of S$_{20}%
$. The $E-k$ dispersion of the $4f$ electron is much more weak for S$_{0}$,
almost forming a flat band at $-4.1$ eV. Since there is $d-f$ hybridization on
each Ce site mediated by I$^{-}$ ions, the $5d$ electronic states are slightly
affected by the $4f$ electronic states as well. The energy gaps of S$_{0}$ and
S$_{20}$, which are caused by the spin splitting of the $5d$ electrons, are
$386$ meV and $336$ meV, respectively. In fact, there exists another
degenerate ground state S$_{0}^{+}$, whose $4f$ electronic state is (0.66, 0,
0, 0.74, 0, 0, 0). This state is recognized to show exactly the same magnetic
properties as S$_{0}$. The detailed discussion about S$_{0}^{+}$ is given in
Section F of the Supporting Information. \cite{SI} The spin densities and
electronic structures of all the other metastable states are also shown in
Section F of the Supporting Information. \cite{SI}

\section{Electronic orbital energy in the RTP crystal field}

In order to have a better understanding of the ground state of FM monolayer
CeI$_{2}$, we calculate the electronic orbital energies of the single $4f$
electron when occupying different orbitals in the RTP crystal field. This
so-called orbital energy is originated from pure Coulomb interactions. As
shown in Fig. 3(a), the single $4f$ electron is located at $O$, with six
surrounding anions located at $R_{i}$ ($x_{i}$, $y_{i}$, $z_{i}$)
($i=1,2,\cdots,6$) in the Cartesian coordinate system. The height of the RTP
is denoted as $l_{0}$ and the edge length of the bottom surface is denoted as
$a_{0}$. $l_{0}$ and $a_{0}$ are taken to be $4.24$ \AA \ and $4.28$ \AA ,
respectively, which are based on the relaxed monolayer CeI$_{2}$ by DFT
calculations. The orbital energy can be expressed as
\begin{equation}
\left\langle E_{4f}\right\rangle =\left\langle \psi_{4f}\right\vert
V_{\text{CF}}\left\vert \psi_{4f}\right\rangle ,
\end{equation}
where $\psi_{4f}$ is the wave function of the single $4f$ electron and
$V_{\text{CF}}$ is the crystal field potential energy with a from of
\begin{equation}
V_{\text{CF}}=\sum_{i=1}^{6}\frac{1}{\sqrt{\left(  x-x_{i}\right)
^{2}+\left(  y-y_{i}\right)  ^{2}+\left(  z-z_{i}\right)  ^{2}}},
\end{equation}
in which the constant term has been set to be unity for simplicity in the
calculations. The charge densities of the six I$^{-}$ ions are treated as
point charges. Since the single $4f$ electron of CeI$_{2}$ is localized, the
numerical integration for Eq. (10) is calculated within a integral area
$\Omega$:%
\begin{equation}
E_{4f}=\int_{\Omega}V_{\text{CF}}\left(  x,y,z\right)  \left\vert \psi
_{4f}\left(  x,y,z\right)  \right\vert ^{2}dxdydz,
\end{equation}
where $\Omega$ denotes the spherical area $\sqrt{x^{2}+y^{2}+z^{2}}\leq R$.
The integral radius $R$ is set to be $1.2$ times of the Wigner-Seitz radius of
Ce to reasonably describe both the local character and the $f\mathtt{-}p$
hybridization of the $4f$ electron.

The orbital energies of all the identified $4f^{1}$ electronic states listed
in Table I are computed. As shown in Fig. 3 (b), the $4f$ electronic state of
S$_{0}$ has the lowest orbital energy in the simulated RTP crystal field
environment of the relaxed monolayer CeI$_{2}$. Now besides the largest value
of $J^{site}$, the RTP crystal field in monolayer CeI$_{2}$ also helps to
stabilize the ground state S$_{0}$ by attaching the lowest $4f$ electronic
orbital energy. In the RTP crystal field with D$_{^{3h}}$ point group
symmetry, the original seven-fold $4f$ orbitals split into three
non-degenerate levels $A_{1}^{\prime}$, $A_{2}^{\prime}$, $A_{2}^{\prime
\prime}$\ and a pair of two-fold levels $E^{\prime}$ and $E^{\prime\prime}$.
The degeneracy of the symmetry-protected orbitals is well-reflected in Fig.
3(b). The $4f$ orbital states are $f_{xyz}$, $f_{z(x^{2}-y^{2})}$, and
$-0.92f_{xyz}+0.4f_{z(x^{2}-y^{2})}$ for S$_{15}$, S$_{16}$ and S$_{17}$,
respectively. As $f_{xyz}$ and $f_{z(x^{2}-y^{2})}$ are belong to degenerate
$E^{\prime\prime}$, the numerical results of crystal orbital energy are the
same for these two orbitals along with their linear combination. The similar
degeneracy occurs in S$_{4}$ and S$_{5}$, or S$_{12}$ and S$_{13}$, which
originates from the degenerate $E^{\prime}$: $f_{yz^{2}}$ and $f_{xz^{2}}$.
What's more, the degenerate $4f$ crystal field orbitals also access to exactly
equal on-site $d-f$ exchange energy and total energy for monolayer CeI$_{2}$
referring to Table I. This feature reflects the strong constrain from orbital
symmetry for the system.

We explore how the orbital energies are influenced by biaxial strain of
monolayer CeI$_{2}$. We examine the seven general $4f$ atomic orbitals and the
$4f$ electronic orbital of S$_{0}$, as shown in Fig. 4 (c). The degeneracy of
two groups of orbitals $E^{\prime}$: $f_{yz^{2}}$\&$f_{xz^{2}}$ and
$E^{\prime\prime}$: $f_{xyz}$\&$f_{z(x^{2}-y^{2})}$ are maintained under
biaxial strain, since biaxial strain does not change the D$_{3h}$ point group
symmetry of monolayer CeI$_{2}$. The $4f$ electronic orbital of S$_{0}$:
$0.66f_{y(3x^{2}-y^{2})}-0.74f_{z^{3}}$ keeps its lowest orbital energy until
the compressive strain reaches 4\% of the lattice parameter. Fig.3 (c) reveals
the change of relative energies among different $4f$ atomic orbitals caused by
strain of the crystal structure. Although the $4f$ orbital energy itself does
not dominate the local-electron ground state of monolayer CeI$_{2}$, we expect
potential strain-caused phase transition of local electronic state in magnetic
$3d$ transition systems, where the crystal field plays a more important role
than $4f$ electronic systems.

Since OMC is applied, the DFT total energy of the system with the $4f$
electron occupying arbitrary orbital is able to be gained as long as the
corresponding $4f$ electronic state is metastable. Due to the high symmetry of
the crystal field in monolayer CeI$_{2}$, the seven general $4f$ atomic
orbitals are almost metastable. The corresponding DFT total energies are
excerpted from Table I into Fig. 3 (d). Fig.3 (d) shows the feasibility of
simulating crystal field excitation by using DFT plus OMC for crystals. This
approach includes electronic interactions from the whole crystal field
environment, such as non-nearest ligand field and orbital characters of the
ligands, which is not that of point charges like in a simple crystal field
model. Thus, the simulated crystal field excitation energy is expected to
agree with experiments better than model calculations.

So far three kinds of electronic interactions have been verified to have
influence on the FM ground state of monolayer CeI$_{2}$ in this paper. They
are inter-site exchange interaction, on-site $d-f$ exchange interaction and
$4f$ crystal field Coulomb interaction. In fact, it is natural to take the
crystal field energy of the $5d$ electron into account. However, with both
local and itinerant characters, the simulations for the $5d$ electron are
complicated and are beyond our study. A qualitative discussion about the $5d$
electronic states in monolayer CeI$_{2}$ is given in Section E of the
supportting information.

\section{The magnetic GS with strong SOC}

In the more realistic situation, the electronic state of the localized
4\textit{f} electron in monolayer CeI$_{2}$ is not only decided by electronic
interactions, but also decided by strong SOC. Thus, the acquirement of the
magnetic GS requires fully unconstrained noncollinear self-consistent
calculations \cite{Worden} with SOC included in the DFT+U energy functional.
Here OMC is once again applied to exclude metastable states. With SOC being
considered, the magnetic ground state of monolayer CeI$_{2}$ still favors FM
order, but the magnetic moments are with components of both $y$ and $z$
directions, as shown in Fig. 4(a). The easy axis of magnetization is coupled
with the crystal field, which makes the magnetic moments point toward certain
directions of the crystal structure. The ratio of the off-plane ($z$) and
in-plane ($y$) components of the total magnetic moment is $0.65:1$. The net
magnetic moment for each Ce is about $1.92$ $\mu_{B}$ which is contributed by
both spin magnetic moment and orbital moment.

The DFT calculations also recommend a 12-fold degenerate magnetic GS for
monolayer CeI$_{2}$ with SOC included, as displayed in Fig. 5(b). All
thedegenerate states can be related to the FM GS shown in Fig. 4(a) by the
symmetry operations of D$_{3h}$ point group, i.e. \{$E$, $2C_{3}$,
$3C_{2}^{\prime}$, $\sigma_{h}$, $2S_{3}$ and $3\sigma_{v}$\}. Due to SOC, the
symmetry operations act on both electronic orbitals and spins. Thus these
degenerate states can be clearly distinguished by the different orientations
of magnetization. Except for the original FM GS, the $C_{3}$ and
$C_{2}^{\prime}$ rotation operations give birth to another five degenerate
states with identical chirality of the electronic orbital. This identical
chirality is marked by \textquotedblleft$+$\textquotedblright\ in Fig. 5(b).
The $\sigma_{v}$ inversion operations then double the number of degenerate
states, and produce six more states with opposite chirality of the electronic
orbital. This opposite chirality is marked by \textquotedblleft$-$%
\textquotedblright\ in Fig. 5(b). We expect this 12-fold degenerate magnetic
GS of monolayer CeI$_{2}$ could be experimentally verified by magnetic measurements.

The SOC bands of monolayer CeI$_{2}$ is shown in Fig. 4(c). The two spins from
$4f$ and $5d$ shells are still parallel to benefit the total energy. No
significant and qualitative changes are detected for the electronic structure
near the fermi level when compared with the situation without SOC. In spite of
this, SOC is essential to reasonably describe the magnetic anisotropy in
monolayer CeI$_{2}$. As shown in Fig. 4(d), with the strong enough external
magnetic field along $x$, $y$, and $z$ direction, respectively, the total
energies of the corresponding magnetized states are $5.53$ meV, $8.60$ meV,
and $0.78$ meV higher per Ce than the magnetic GS. These energy differences
are one order larger than the situation\ that SOC is treated as a
perturbation. The variation of the electronic orbital contributes to these
energy differences, in addition to the SOC energy introduced by Eq. (9). Hence
the conclusion that monolayer CeI$_{2}$ is an isotropic in-plane FM material
is no longer satisfied when getting away from the framework of weak SOC which
is discussed in Section IV.

The fully off-plane ($z$ direction) FM state is concerned due to its potential
application in spintronic devices. A comparison between this FM metastable
state and the FM GS is shown in Table II. The included angle of the magnetic
moments from the two FM states is 57${}^{\circ}$. The fully off-plane FM state
is 0.78 meV higher per Ce in total energy that the FM GS. Apparently an
appropriate external magnetic field can help to achieve saturated
magnetization in the off-plane direction from the GS. We also expect other
measures that can tune the magnetic characters of monolayer CeI$_{2}$ such as
electric field or strain.

\begin{table}[ptbh]
\caption{The comparison between the FM GS ($G_{\text{FM}}^{0}$) and the fully
off-plane FM metastable state ($G_{\text{FM}}^{z}$) of monolayer CeI$_{2}$
with SOC included. The $2^{\text{nd}}$ column shows the magnetic moment
vectors $\vec{S}$ of the two states. The $3^{\text{rd}}$ column shows the
relative energy $\Delta E$ (in unit of meV) of the two states. The
$4^{\text{th }}$column shows the specific representation of the 4$f^{1}$
electronic states with $f_{y(3x^{2}-y^{2})}$, $f_{xyz}$, $f_{yz^{2}}$,
$f_{z^{3}}$, $f_{xz^{2}}$, $f_{z(x^{2}-y^{2})}$, and $f_{x(x^{2}-3y^{2})}$ as
bases. $\uparrow$ and $\downarrow$ denote the status of spin.}%
\label{tableII}%
\centering
\begin{tabular*}
{\hsize}[c]{cccc}\hline\hline
& $\vec{S}$ & $\Delta E$ (meV) & $4f^{1}$ electronic states\\\hline
\multirow{2}{*}{$G_{\text{FM}}^{z}$} & \multirow{2}{*}{$(0,0,1.92)$} &
\multirow{2}{*}{$0.78$} & $(0.61,0,0,-0.72,0,0,0.267i)^{\uparrow}$\\
&  &  & $+(0,0,-0.11i,0,-0.11,0,0)^{\downarrow}$\\\hline
\multirow{2}{*}{$G_{\text{FM}}^{0}$} & \multirow{2}{*}{$(0,1.61,1.06)$} &
\multirow{2}{*}{$0$} & $(0.57,0,-0.15i,0.6,0,0,0.19i)^{\uparrow}$\\
&  &  & $+(-0.23+0.15i,0,0.14i,-0.3+0.18i,0.1,0,0)^{\downarrow}$\\\hline\hline
\end{tabular*}
\end{table}

\section{Conclusions}

In this work, the ab-initio DFT+U calculations combined with OMC are performed
to investigate the magnetic GS of monolayer CeI$_{2}$. The FM GS and FM
metastable states are identified. It is shown that the metastable states have
different magnetic properties with the magnetic GS. We recommend that
something ought to be done to prevent the metastable states from producing
errors when predicting the magnetic properties of such local-electron systems
by DFT calculations. The $T_{\text{C}}$ of monolayer CeI$_{2}$ is simulated to
be about $128$ K based on the magnetic ground state, not above room
temperature as mentioned before. The calculations of on-site $d-f$ exchange
parameter and $4f$ electronic orbital energy imply that the GS of monolayer
CeI$_{2}$ with FM order requires the harmony of multiple types of
electron-electron interactions to minimize the total energy. With SOC
included, the easy axis of magnetization is found to be coupled with the
crystal structure. The FM GS with both in-plane and off-plane components of
magnetic moments, and a fully off-plane FM metastable state of monolayer
CeI$_{2}$ are confirmed to be close in total energy. Our work shows it is
necessary to absorb SOC into the energy functional to allow SOC to affect the
electronic state of the local electrons during the self-consistent
calculations. Only in this way could the reliable magnetic GS of monolayer
CeI$_{2}$ be obtained and the magnetic anisotropy be reliably described.

For other systems with local-electron magnetism, the identification of the
local electronic states are also significant. However, the phenomenon of
trapping into metastable states during the calculations is quite common and it
causes confusions if no extra information is given. Here it is recommended to
give the specific representation of the computed local electronic state in
one's research. This action helps to enhance the repeatability and
normalization of the computational studies for magnetic materials. (All the
representations of the used starting points and corresponding final states in
our work have been shown in Section G of the Supporting Information. \cite{SI})

\begin{acknowledgments}
We thank Doctor Yuan Hong from Institute of Applied Physics and Computational
Mathematics for his useful discussions with us. We thank our referees for the
time they paid on our manuscript. We are grateful for their helpful comments
and suggestions. This work was supported by the National Natural Science
Foundation of China (Grant No. 12175023, and No. 12104034).
\end{acknowledgments}

\begin{figure}[ptb]
	\begin{center}
		\includegraphics[width=1.0\linewidth]{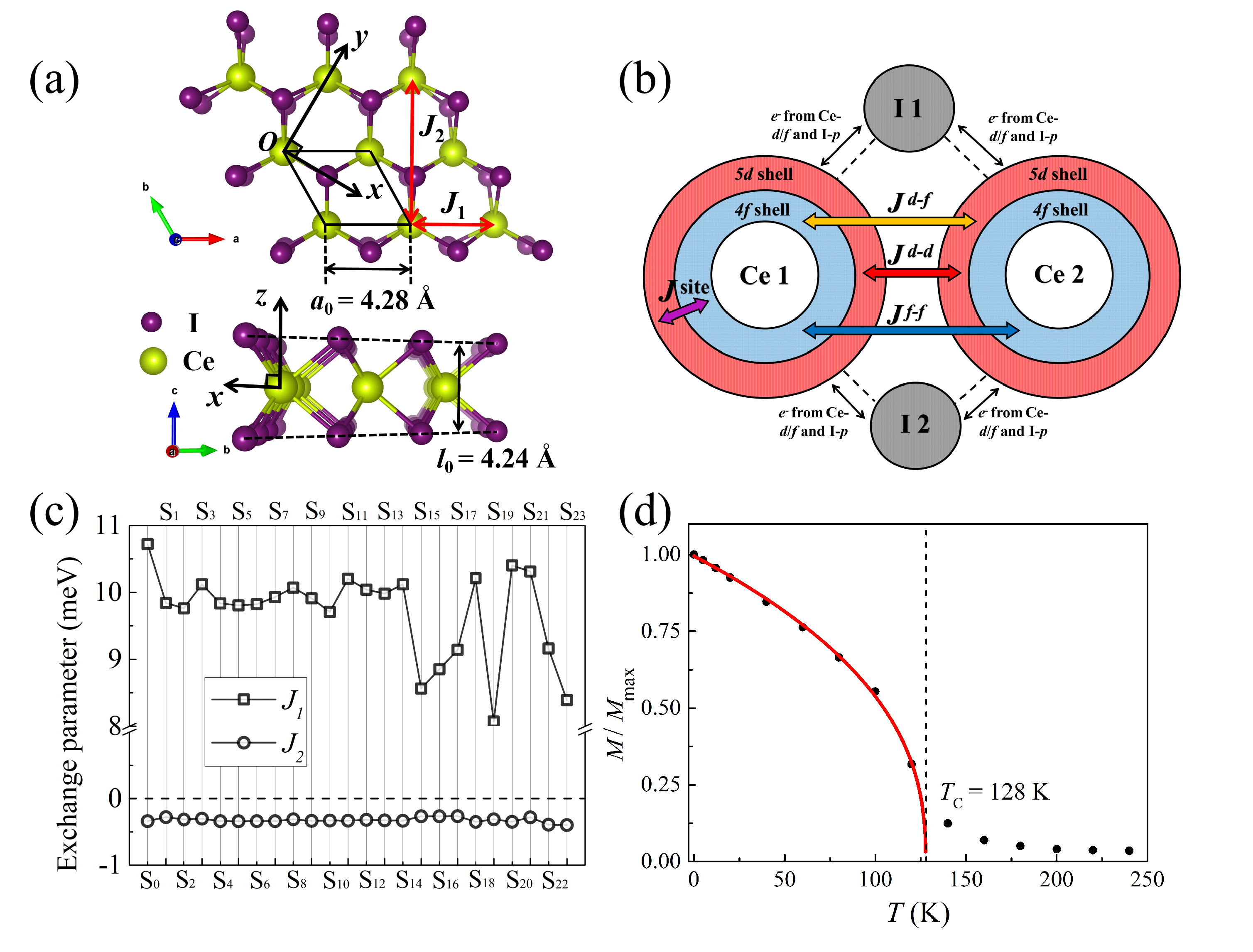}
	\end{center}
	\caption{(Color online) (a) The optimized crystal structure of monolayer
		CeI$_{2}$. $a_{0}$ and $l_{0}$ are the lattice parameter and thickness of
		monolayer CeI$_{2}$. (b) The components ($J^{d-d}$, $J^{d-f}$, and $J^{f-f}$)
		of the first nearest inter-site exchange parameter and the on-site $d-f$
		exchange parameter ($J^{\text{site}}$) in monolayer CeI$_{2}$. (c) The values
		of first nearest exchange parameter $J_{1}$ and second nearest exchange
		parameter $J_{2}$ for the GS and metastable states. (d) The temperature
		dependent magnetization intensity of monolayer CeI$_{2}$ with a quenching at
		$128$ K predicted by Monte-Carlo simulations.}%
	\label{fig1}%
\end{figure}

\begin{figure}[ptb]
	\begin{center}
		\includegraphics[width=1.0\linewidth]{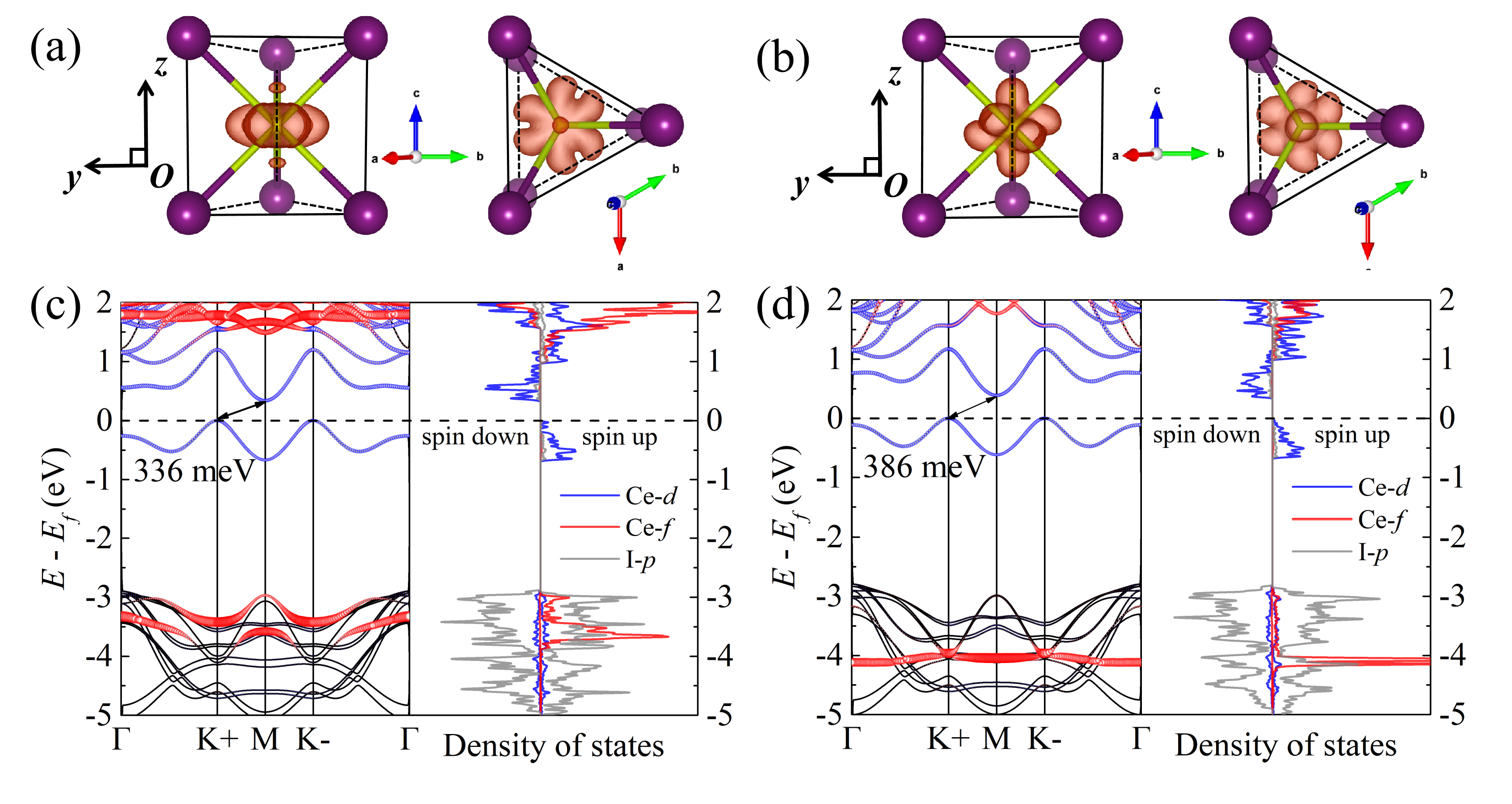}
	\end{center}
	\caption{(Color online) The isosurfaces of spin densities for (a): S$_{20}$
		and (b): S$_{0}$. The energy bands and density of states of (c): S$_{20}$ and
		(d): S$_{0}$. The contributions of Ce-$d$ and Ce-$f$ orbitals to the bands are
		represented by the blue and red bubbles, respectively.}%
	\label{fig2}%
\end{figure}

\begin{figure}[ptb]
	\begin{center}
		\includegraphics[width=0.95\linewidth]{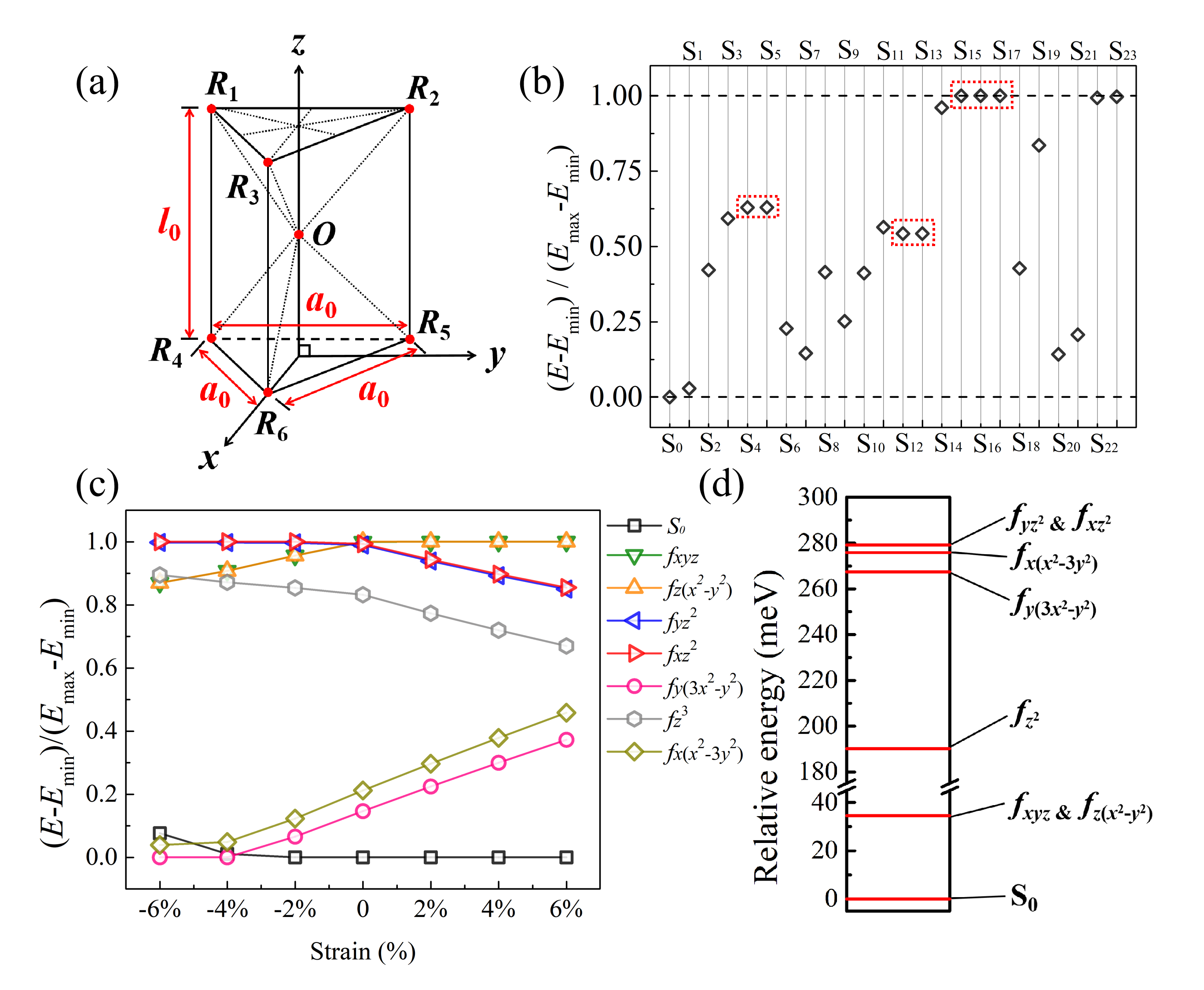}
	\end{center}
	\caption{(Color online) (a) The RTP crystal field model in monolayer CeI$_{2}$
		with the first nearest I$^{-}$ ions being considered only. (b) The relative
		$4f$ electronic orbital energies of the $24$ identified FM states in the RTP
		crystal field environment of relaxed monolayer CeI$_{2}$. The lowest orbital
		energy is set to $0$. The degenerate orbitals are marked by a red dashed box.
		(c) The $4f$ orbital energies of S$_{0}$: $0.66f_{y(3x^{2}-y^{2}%
			)}-0.74f_{z^{3}}$ and the seven general $4f$ orbitals in the RTP crystal
		fields environment of biaxial-strain monolayer CeI$_{2}$. The corresponding
		RTP crystal fields are simulated in the range of 6\% compressive strain to 6\%
		tensile strain of the lattice parameter. (d) The relative DFT total energies
		of monolayer CeI$_{2}$ with the $4f$ electron occupying different orbitals
		which are discussed in (c).}%
	\label{fig3}%
\end{figure}

\begin{figure}[ptb]
	\begin{center}
		\includegraphics[width=1.0\linewidth]{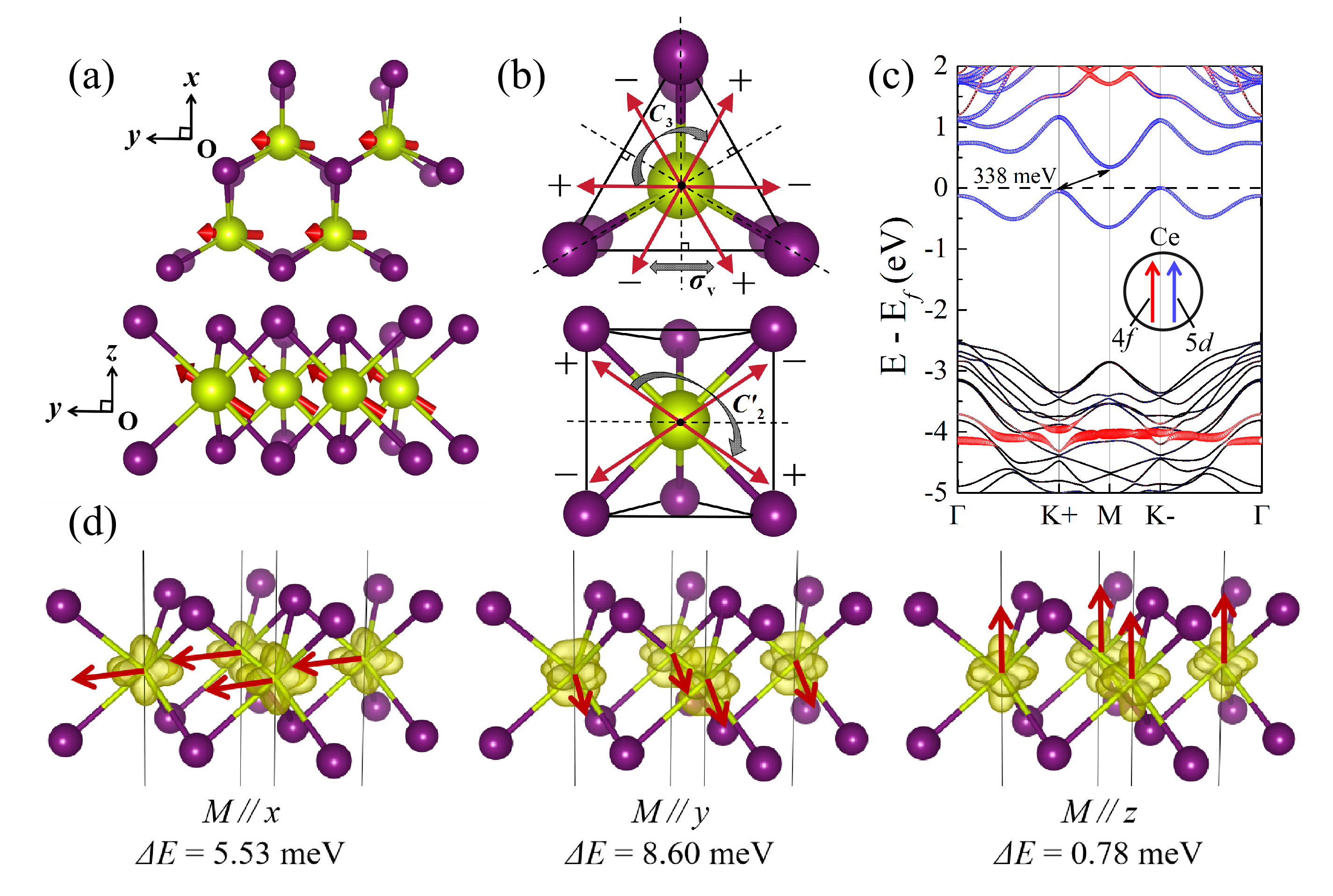}
	\end{center}
	\caption{(Color online) (a) The FM structure of the GS of monolayer CeI$_{2}$
		with SOC included. The component of $x$ direction for the magnetic moment is
		strictly equal to $0$. (b) The 12-fold degenerate magnetic GS of monolayer
		CeI$_{2}$. The red arrows represent the orientations of magnetization of
		different degenerate states. (c) The SOC energy bands of the GS of monolayer
		CeI$_{2}$. The contributions of Ce-$d$ and Ce-$f$ orbitals to the bands are
		represented by the blue and red bubbles, respectively. (d) The magnetization
		densities of the FM states with the external magnetic field along $x$, $y$,
		and $z$ directions. The red arrows mark the directions of the magnetic moment
		vectors. $\Delta E$ is the total energy of the magnetized state relative to
		the FM GS.}%
	\label{fig4}%
\end{figure}

\end{document}